\documentclass[12pt]{article}
 \usepackage{epsfig}
 \def\be{\begin{equation}}
 \def\ee{\end{equation}}
 \def\bea{\begin{eqnarray}}
 \def\eea{\end{eqnarray}}
 \usepackage{graphicx}

 \catcode`\@=11
 \def\lsim{\mathrel{\mathpalette\@versim<}}
 \def\gsim{\mathrel{\mathpalette\@versim>}}
 \def\@versim#1#2{\vcenter{\offinterlineskip
 \ialign{$\m@th#1\hfil##\hfil$\crcr#2\crcr\sim\crcr } }}
 \catcode`\@=12

 \catcode`@=12
\begin{document}
 \thispagestyle{empty}
 \begin{flushright}
 UCRHEP-T561\\
 January 2016\
 \end{flushright}
 \vspace{0.6in}
 \begin{center}
 {\LARGE \bf Diphoton Revelation of the Utilitarian\\
 Supersymmetric Standard Model\\}
 \vspace{1.2in}
 {\bf Ernest Ma\\}
 \vspace{0.2in}
 {\sl Physics \& Astronomy Department and Graduate Division,\\ 
 University of California, Riverside, California 92521, USA\\}
 \vspace{0.1in}
 {\sl HKUST Jockey Club Institute for Advanced Study,\\ 
 Hong Kong University of Science and Technology, Hong Kong, China\\}

 \end{center}
 \vspace{1.2in}

\begin{abstract}\
In 2002, I proposed a unique $U(1)$ extension of the supersymmetric 
standard model which has no $\mu$ term and conserves baryon number and 
lepton number separately and automatically.  This model, \underline{without 
any change}, has all the necessary and sufficient ingredients to explain 
the possible 750 GeV diphoton excess, observed recently by the ATLAS 
Collaboration at the Large Hadron Collider (LHC).  It is associated with 
the superfield which replaces the $\mu$ parameter.  If confirmed and 
supported by subsequent data, it may even be considered as the first evidence 
for supersymmetry.
\end{abstract}

 \newpage
 \baselineskip 24pt

Since the recent announcement~\cite{atlas15} by the ATLAS Collaboration 
at the Large Hadron Collider (LHC) of a diphoton excess around 750 GeV, 
numerous papers~\cite{many} have appeared explaining its presence or 
discussing its implications.  In this short note, I simply point out 
that an explicit model I proposed in 2002~\cite{m02} has exactly all 
the necessary and sufficient particles and interactions for this purpose. 
They were of course there for solving other issues in particle physics.  
However, the observed diphoton excess may well be a first revelation of 
this model, including its connection to dark matter.

This 2002 model extends the supersymmetric standard model by a new $U(1)_X$ 
gauge symmetry.  It replaces the $\mu$ term with a singlet scalar 
superfield which also couples to heavy color-triplet superfields which 
are electroweak singlets.  The latter are not {\it ad hoc} inventions, 
but are necessary for the cancellation of axial-vector anomalies.  
It was shown in Ref.~\cite{m02} how this was accomplished by the 
remarkable \underline{exact factorization} of the 
sum of eleven cubic terms, resulting in two generic classes of solutions. 
Both are able to enforce the conservation of baryon number and lepton 
number up to dimension-five terms.  As such, both the scalar singlet and 
the vectorlike quarks are essential predictions of this 2002 model. 
They are thus naturally suited for explaining the observed diphoton excess.
In 2010~\cite{m10}, I discussed a specific version which will be used here 
as well to illustrate my point.

\noindent \underline{\it Model}~:~ Consider the gauge group $SU(3)_C \times
SU(2)_L \times U(1)_Y \times U(1)_X$ with the particle content of 
Ref.~\cite{m02}.  For $n_1=0$ in Solution (A), the various superfields 
transform as shown in Table 1.  There are three copies of 
$Q,u^c,d^c,L,e^c,N^c,S_1,S_2$; two copies of $U,U^c,S_3$; and one copy of 
$\phi_1,\phi_2,D,D^c$.
\begin{table}[htb]
\caption{Particle content of proposed model.}
\begin{center}
\begin{tabular}{|c|c|c|c|c|}
\hline
Superfield & $SU(3)_C$ & $SU(2)_L$ & $U(1)_Y$ & $U(1)_X$ \\
\hline
$Q = (u,d)$ & 3 & 2 & 1/6 & 0 \\
$u^c$ & $3^*$ & 1 & $-2/3$ & 3/2 \\
$d^c$ & $3^*$ & 1 & 1/3 & 3/2 \\
\hline
$L = (\nu,e)$ & 1 & 2 & $-1/2$ & 1 \\
$e^c$ & 1 & 1 & 1 & 1/2 \\
$N^c$ & 1 & 1 & 0 & 1/2 \\  
\hline
$\phi_1$ & 1 & 2 & $-1/2$ & $-3/2$ \\
$\phi_2$ & 1 & 2 & 1/2 & $-3/2$ \\
$S_1$ & 1 & 1 & 0 & $-1$ \\
$S_2$ & 1 & 1 & 0 & $-2$ \\
$S_3$ & 1 & 1 & 0 & 3 \\
\hline
$U$ & 3 & 1 & 2/3 & $-2$ \\
$D$ & 3 & 1 & $-1/3$ & $-2$ \\
$U^c$ & $3^*$ & 1 & $-2/3$ & $-1$ \\
$D^c$ & $3^*$ & 1 & 1/3 & $-1$ \\
\hline
\end{tabular}
\end{center}
\end{table}
The only allowed terms of the superpotential are thus trilinear, i.e.
\begin{eqnarray}
&& Q u^c \phi_2, ~~~ Q d^c \phi_1, ~~~ L e^c \phi_1, ~~~ L N^c \phi_2, ~~~ 
S_3 \phi_1 \phi_2, ~~~ N^c N^c S_1, \\ 
&& S_3 U U^c, ~~~ S_3 D D^c, ~~~ u^c N^c U, ~~~ u^c e^c D, ~~~ d^c N^c D, ~~~ 
Q L D^c, ~~~ S_1 S_2 S_3.
\end{eqnarray}
The absence of any bilinear term means that all masses come from soft 
supersymmetry breaking, thus explaining why the $U(1)_X$ and electroweak 
symmetry breaking scales are not far from that of supersymmetry breaking. 
As $S_{1,2,3}$ acquire nonzero vacuum expectation values (VEVs), the exotic 
$(U,U^c)$ and $(D,D^c)$ fermions obtain Dirac masses from $\langle S_3 
\rangle$, which also generates the $\mu$ term.  The singlet $N^c$ fermion 
gets a large Majorana mass from $\langle S_1 \rangle$, so that the neutrino 
$\nu$ gets a small seesaw mass in the usual way. The singlet $S_{1,2,3}$ 
fermions themselves get Majorana masses from their scalar counterparts 
$\langle S_{1,2,3} \rangle$ through the $S_1 S_2 S_3$ terms.  The only 
massless fields left are the usual quarks and leptons. They then become 
massive as $\phi^0_{1,2}$ acquire VEVs, as in the minimal supersymmetric 
standard model (MSSM).

Because of $U(1)_X$, the structure of the superpotential conserves both 
$B$ and $(-1)^L$, with $B=1/3$ for $Q,U,D$, and $B=-1/3$ for $u^c,d^c,U^c,D^c$; 
$(-1)^L$ odd for $L,e^c,N^c,U,U^c,D,D^c$, and even for all others. Hence 
the exotic $U,U^c,D,D^c$ scalars are leptoquarks and decay into ordinary quarks 
and leptons.  The $R$ parity of the MSSM is defined here in the same way, 
i.e. $R \equiv (-)^{2j+3B+L}$, and is conserved.  Note also that the 
quadrilinear terms $QQQL$ and $u^c u^c d^c e^c$ (allowed in the MSSM) as 
well as $u^c d^c d^c N^c$ are forbidden by $U(1)_X$.  Proton decay is thus 
strongly suppressed.  It may proceed through the quintilinear term 
$QQQL S_1$ as the $S_1$ fields acquire VEVs, but this is a dimension-six 
term in the effective Lagrangian, which is suppressed by two powers 
of a very large mass, say the Planck mass, and may safely be allowed.

\noindent \underline{\it Gauge sector}~:~ The new $Z_X$ gauge boson of this 
model becomes massive through $\langle S_{1,2,3} \rangle = u_{1,2,3}$, whereas 
$\langle \phi^0_{1,2} \rangle = v_{1,2}$ contribute to both $Z$ and $Z_X$. 
The resulting $2 \times 2$ mass-squared matrix is given by~\cite{km97}
\begin{equation}
{\cal M}^2_{Z,Z_X} = \pmatrix{(1/2)g_Z^2(v_1^2+v_2^2) & (3/2)g_Z g_X 
(v_2^2-v_1^2) \cr (3/2)g_Z g_X (v_2^2-v_1^2) & 2g_X^2 [u_1^2 + 4 u_2^2 + 
9 u_3^2 + (9/4)(v_1^2 + v_2^2)]}.
\end{equation}
Since precision electroweak measurements require $Z-Z_X$ mixing to be very 
small~\cite{elmp09}, $v_1 = v_2$, i.e. $\tan \beta = 1$, is preferred. 
With the 2012 discovery~\cite{atlas12,cms12} of the 125 GeV 
particle, and identified as the one Higgs boson $h$ responsible for 
electroweak symmetry breaking, $\tan \beta =1$ is not compatible with the 
MSSM, but is perfectly consistent here, as shown already in Ref.~\cite{m10}.

Consider the decay of $Z_X$ to the usual quarks and leptons.  Each fermionic 
partial width is given by
\begin{equation}
\Gamma(Z_X \to \bar{f} f) = {g_X^2 M_{Z_X} \over 24 \pi} [c_L^2 + c_R^2],
\end{equation}
where $c_{L,R}$ can be read off under $U(1)_X$ from Table 1.  Thus
\begin{equation}
{\Gamma(Z_X \to \bar{t} t) \over \Gamma(Z_X \to \mu^+ \mu^-)} = 
{\Gamma(Z_X \to \bar{b} b) \over \Gamma(Z_X \to \mu^+ \mu^-)} = {27 \over 5}.
\end{equation}
This will serve to distinguish it from other $Z'$ models~\cite{gm08}.

\noindent \underline{\it Relevance to the observed diphoton excess}~:~
In this model, other than the addition of $N^c$ for seesaw neutrino masses, 
the only new particles are $U, U^c, D, D^c$ and $S_{1,2,3}$, which are 
exactly the ingredients needed to explain the diphoton excess at the LHC. 
The allowed $S_3 U U^c$ and $S_3 D D^c$ couplings enable the one-loop gluon 
production of $S_3$ in analogy to that of $h$.  
\begin{figure}[htb]
\vspace*{-3cm}
\hspace*{-3cm}
\includegraphics[scale=1.0]{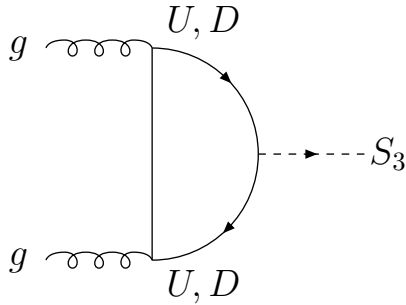}
\vspace*{-21.5cm}
\caption{One-loop production of $S_3$ by gluon fusion.}
\end{figure}
The one-loop decay of 
$S_3$ to two photons comes from these couplings as well as 
$S_3 \phi_1 \phi_2$. 
\begin{figure}[htb]
\vspace*{-3cm}
\hspace*{-3cm}
\includegraphics[scale=1.0]{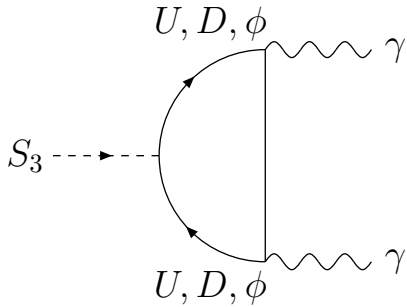}
\vspace*{-21.5cm}
\caption{One-loop decay of $S_3$ to two photons.}
\end{figure}
In addition, the direct $S_1 S_2 S_3$ couplings 
enable the decay of $S_3$ to other final states, including those of the 
dark sector, which contribute to its total width.  The fact that the 
exotic $U,U^c,D,D^c$ scalars are leptoquarks is also very useful for 
understanding~\cite{bn15} other possible LHC flavor anomalies.  In a 
nutshell, a desirable comprehensive picture of possible new physics 
beyond the standard model is encapsulated by this existing model.
Note that there are two copies of $S_3$, so it is possible that the 
750 GeV particle is a pseudoscalar.  This has the advantage that it 
does not mix with the 125 GeV Higgs scalar.

The phenomenological analysis to support the above claim is similar to 
all other recent proposals using this mechanism, and follows closely 
that of Ref.~\cite{bn15}, including the necessary addition of the 
contribution of the color-singlet charged $\phi$ fermions to increase the 
$S_3 \to \gamma \gamma$ rate.  Note that this is predicted by the model 
and not an {\it ad hoc} invention.  The possibility of 
$S_3 \to S_1 S_2$ here could also enhance its invisible width and serves 
as a link to dark matter.

\noindent \underline{\it Predictions}~:~
Since $S_3$ couples to leptoquarks, the $S_3 \to l_i^+ l_j^-$ decay must 
occur at some level.  As such, $S_3 \to e^+ \mu^-$ would be a very 
distinct signature at the LHC.  Its branching fraction depends on unknown 
Yukawa couplings which need not be very small.  Similarly, the $S_3$ 
couplings to $\phi_1 \phi_2$ as well as leptoquarks imply decays to $ZZ$ and 
$Z \gamma$ with rates comparable to $\gamma \gamma$.

\noindent \underline{\it Conclusion}~:~
The utilitarian supersymmetric $U(1)_X$ gauge extension of the Standard Model 
of particle interactions proposed 14 years ago~\cite{m02} allows for two 
classes of anomaly-free models which have no $\mu$ term and conserve baryon 
number and lepton number automatically.  A simple version~\cite{m10} with 
leptoquark superfields is especially interesting because of existing LHC 
flavor anomalies.

The new $Z_X$ gauge boson of this model has specified couplings to quarks and 
leptons which are distinct from other gauge extensions and may be tested at 
the LHC.  On the other hand, a hint may already be discovered with the 
recent announcement by ATLAS of a diphoton excess at around 750 GeV.  It 
may well be the revelation of the singlet scalar (or pseudoscalar) $S_3$ 
predicted by this 
model which also predicts that there should be singlet leptoquarks and 
other particles that $S_3$ must couple to.  Consequently, gluon fusion 
will produce $S_3$ which will then decay to two photons together with 
other particles, including those of the dark sector.  This scenario 
explains the observed diphoton excess, all within the context of the  
original model, and not an invention after the fact.

Most importantly, since $S_3$ replaces the $\mu$ parameter, its identification 
with the 750 GeV excess implies the existence of supersymmetry.  If confirmed 
and supported by subsequent data, it may even be considered in retrospect 
as the first evidence for the long-sought existence of supersymmetry.

\noindent \underline{\it Acknowledgement}~:~
This work was supported in part by the U.~S.~Department of Energy Grant 
No. DE-SC0008541.

\bibliographystyle{unsrt}

\end{document}